\documentclass[aps,prc,twocolumn,showpacs,floatfix,nofootinbib,preprintnumbers,superscriptaddress,amsmath,amssymb]{revtex4-1}

\usepackage{graphicx}
\usepackage{dcolumn}
\usepackage{bm}
\usepackage{hyperref}
\usepackage[mathlines]{lineno}

\newcommand{\hfbax}{\sc hfb-ax}
\newcommand{\hfbtho}{\sc hfbtho}

\begin{document}

\title{ Emergent soft monopole modes in weakly-bound deformed nuclei}

\author{J.C. Pei}
\affiliation{State Key Laboratory of Nuclear
Physics and Technology, School of Physics, Peking University,  Beijing 100871, China}

\author{M. Kortelainen}
\affiliation{Department of Physics, P.O. Box 35 (YFL), University of Jyv\"askyl\"a, FI-40014 Jyv\"askyl\"a, Finland}
\affiliation{Helsinki Institute of Physics, P.O. Box 64, FI-00014 University of Helsinki, Finland}

\author{Y.N. Zhang}
\affiliation{State Key Laboratory of Nuclear
Physics and Technology, School of Physics, Peking University,  Beijing 100871, China}

\author{F.R. Xu}
\affiliation{State Key Laboratory of Nuclear
Physics and Technology, School of Physics, Peking University,  Beijing 100871, China}

\begin{abstract}
 Based on the Hartree-Fock-Bogoliubov
solutions in large deformed coordinate spaces, the finite amplitude method for quasiparticle random phase approximation (FAM-QRPA) has been implemented, providing a suitable approach to probe collective excitations of weakly-bound nuclei embedded in the continuum.
The monopole excitation modes in Magnesium isotopes up to the neutron drip line have been studied with the FAM-QRPA framework on
both the coordinate-space and harmonic oscillator basis methods. Enhanced soft monopole strengths and collectivity as a result of weak-binding effects have been unambiguously demonstrated.
\end{abstract}

\pacs{21.10.Gv, 21.10.Re, 21.60.Jz}
\maketitle


Nuclei close to the particle drip lines are weakly bound superfluid quantum systems
and can exhibit exotic threshold phenomena~\cite{halo}, sharing interdisciplinary interests with weakly bound systems such as  multi-quark states, Rydberg atoms and quantum droplets~\cite{quark,halo,droplet}.
Since the discovery of nuclear halos with radioactive beams~\cite{halo-exp}, there have been numerous theoretical developments aiming
at weakly-bound nuclei and their dilute surfaces~\cite{forssen}.
Extensive Hartree-Fock-Bogoliubov (HFB) studies have provided
successful descriptions of continuum couplings and halo features in ground states of weakly-bound nuclei~\cite{continuum-jacek,Mizutori,continuum,schunck,matsuo2009,sgzhou,pei2013}.
On the other hand, excitations in weakly-bound nuclei opened vast possibilities to probe novel collective modes,
as well as continuum effects and components of the effective interaction
that are suppressed in ground states~\cite{yoshida09,paar,dipole}.
To address these issues, along with forthcoming facilities such as FRIB, an accurate and self-consistent treatment of continuum together with pairing correlations, deformations and large spatial extensions is essential.

Among the excited states in weakly-bound nuclei, the emergent soft excitation modes (or pygmy resonances) which correspond to the collective motion between neutron halo/skins and cores, are particularly intriguing. These modes can impact astrophysical neutron capture rates and $r$-process nucleosynthesis. However, the collectivity of observed pygmy resonances, as a crucial verification of coherence, is still under debate~\cite{paar,dipole,reinhard}.
This Rapid Communication is devoted to the low-energy monopole excitations in weakly-bound nuclei, caused  due
to the soft incompressibility of halos, as the dilute nuclear matter has a decreased incompressibility
compared to saturated densities~\cite{khan12}. The low-energy monopole modes indeed
have been predicted, e.g., in the neutron-rich Nickel isotopes (observed very recently in $^{68}$Ni~\cite{vandebrouck}), as a rather non-collective excitation~\cite{nickel}; however, it may hardly be expected in another RPA calculation with a proper treatment of continuum~\cite{hamamoto}.
Besides, the collectivity could be enhanced due to weak-binding effects~\cite{yoshida09}.
Therefore, the emergence of collective soft monopole modes,
as well as the role of continuum contributions with the fully self-consistent continuum quasiparticle random phase approximation (QRPA)
approach, is still an open question.

The standard method to solve the QRPA equation as a matrix form involves tremendous computational costs in deformed cases, and
even more when continuum configurations are included~\cite{terasaki}.
Recently, the developments
of the Finite Amplitude Method (FAM), by Nakatsukasa {\it et. al.}, provided an alternative way to solve the QRPA problem iteratively~\cite{fam07,fam}
rather than  by a direct matrix diagonalization. The FAM-QRPA method provides an efficient way to
study collective excitations and it has been implemented on several HFB approaches, such
as the spherical coordinate-space HFB~\cite{fam}, deformed harmonic oscillator (HO) and transformed HO basis HFB~\cite{hfbtho-fam},
and deformed relativistic Hartree-Bogoliubov method~\cite{rfam}. Recently, FAM-QRPA method has been also applied to the discrete modes~\cite{nobuo}
and $\beta$-decays~\cite{mustonen}.

In this Rapid Communication, we have developed a FAM-QRPA approach based on HFB solutions in large, axially symmetric coordinate-spaces,
to describe excitations in weakly-bound deformed nuclei, which was a great computationally challenge. In deformed weakly-bound nuclei, the subtle interplay
among surface deformations, surface diffuseness, and continuum couplings can result in exotic structures, such as deformed halos.
Therefore theoretical studies of ground state properties and excitations need precise HFB solutions to account these phenomena.
The conventional HFB approach, based on the HO basis may not be sufficient to describe the surface properties of weakly bound systems.
On the other hand, the exact treatment of the continuum in deformed cases, with scattering boundary conditions, is rarely employed~\cite{matsuo2009}.
In this context, the HFB approach in large deformed coordinate-spaces can provide very precise
descriptions of ground states in deformed weakly-bound nuclei, including quasiparticle resonances and dense continuum spectra, and this has been accomplished recently with a hybrid parallel calculation scheme~\cite{pei2013}.
Therefore, the next natural step is to combine the FAM-QRPA method with the deformed large coordinate-space HFB approach,
to realize the deformed continuum QRPA calculations.

The HFB equation is solved by {\hfbax}~\cite{Pei08,pei2011} within a large two-dimensional coordinate space,
based on B-spline techniques for axially symmetric deformed nuclei~\cite{teran}.
The maximum mesh spacing is 0.6\,fm and the order of the B-splines is 12.
For calculations employing large box sizes and small lattice spacings,
the discretized continuum spectra would be very dense, providing good resolutions.
To describe systems in large coordinate spaces, a hybrid MPI+OpenMP parallel programming scheme was implemented to get
converged results within a reasonable time. We note that the {\hfbax} code has
been improved to describe large systems~\cite{pei2013}, since its initial version~\cite{Pei08}.
For the particle-hole interaction channel, the often used Skyrme parameterizations SLy4~\cite{sly4} and
SkM*~\cite{skm} are adopted.
For the particle-particle channel, the density dependent surface pairing interaction
is used~\cite{mix-pairing}.  With a pairing window of 60\,MeV, the pairing strengths are taken as $V_0=500\,{\rm MeV\,fm^3}$ for SLy4 and
$V_0=450\,{\rm MeV\,fm^3}$ for SkM*, so that pairing gaps in stable nuclei can be reasonably reproduced.

Once the HFB solutions are obtained, the quasiparticle wavefunctions and  energies are
recorded. The wavefunctions are represented in a 2D Gauss-Legendre lattice for integral calculations.
Our goal is to solve iteratively the non-linear FAM-QRPA equations~\cite{fam}
\begin{equation}
\begin{array}{c}
\left(E_\mu + E_\nu -\omega\right) X_{\mu \nu}(\omega) +\delta H^{20}_{\mu \nu}(\omega) =-F^{20}_{\mu \nu} ,  \\[4pt]
\left(E_\mu + E_\nu + \omega \right) Y_{\mu \nu}(\omega) +\delta H^{02}_{\mu \nu}(\omega) =-F^{02}_{\mu \nu} ,
\label{FAM}
\end{array}
\end{equation}
where $X_{\mu\nu}(\omega)$ and $Y_{\mu\nu}(\omega)$ are the FAM-QRPA amplitudes;
$\delta H^{20}_{\mu \nu}$ and $\delta H^{02}_{\mu \nu}$ are the induced fields;
$F$ is the external time-dependent field to polarize the system.
Transition strength can be calculated from the $X_{\mu \nu}(\omega)$ and $Y_{\mu \nu}(\omega)$ amplitudes.
The expressions for $H^{20}$ and $H^{02}$ in the HFB approach can be found in Refs.~\cite{fam,rfam}.
We note that time-odd terms of the energy density are very important and have been included.
An imaginary part of the frequency $\omega$ is taken to be 0.5\,MeV for smoothing resonances.
In the coordinate-space approach, the evaluation of $\delta H^{20}$ and $\delta H^{02}$
in terms of a integral of wavefunctions takes the majority of computing time.  The total memory
of wavefunctions and their derivatives takes about 10-20 Gb. For each frequency point $\omega$,
the calculation employs the OpenMP shared memory parallel scheme. For different frequencies,
the MPI distributed parallel scheme is adopted. Consequently, the FAM-QRPA approach can be efficiently
implemented by using this hybrid parallel scheme. The calculations take about 40$\thicksim$80 iterations to converge,
by utilizing the Broyden iteration method (with 30 previous iterations included)~\cite{broyden}, depending on the size of configuration space, i.e., the box size and the quasiparticle energy cutoff.

Although deformed coordinate-space FAM-QRPA
is computationally more expensive compared to the HO basis approach~\cite{hfbtho-fam}, it is much less intensive
than the coordinate-space matrix QRPA approach~\cite{terasaki,yoshida09}. At the moment, deformed
coordinate-space QRPA approaches typically use smaller box sizes which can cause false peaks in the strength function. Compared to a typical QRPA matrix of 160,000 by 160,000 with a box of 20\,fm~\cite{terasaki},
the present FAM-QRPA handles a typical matrix size of 24,000 by 24,000 with a box size of 27\,fm,
implying computational costs can be significantly reduced. Thus this method offers an unique advantage
to study excitations in weakly bound deformed nuclei that require large coordinate spaces.

 \begin{figure}[t]

  \includegraphics[width=0.48\textwidth]{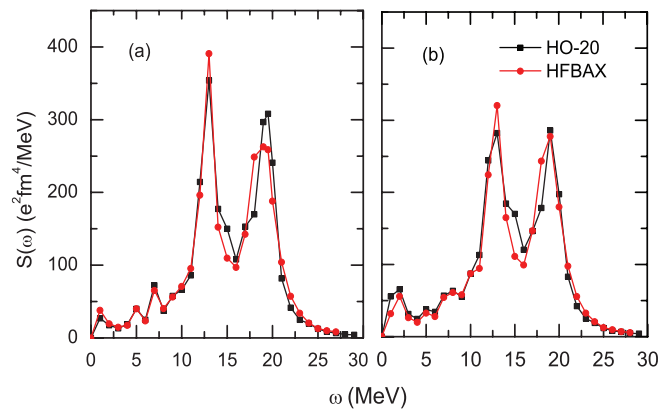}\\
  \caption{(Color online) The isoscalar monopole strength in $^{100}$Zr calculated with the
  FAM-QRPA method based on {\hfbax} and {\hfbtho}, respectively;
  (a) with the volume pairing and (b) with the surface pairing. }
  \label{zr100}
\end{figure}

To verify our implementation of FAM-QRPA, we have calculated the isoscalar (IS) monopole excitation strength function
for $^{100}$Zr, with the SLy4 interaction, which is a well deformed nucleus and has been studied in several earlier works~\cite{hfbtho-fam,yoshida10}.
In Fig.~\ref{zr100}, the calculated transition strengths with different pairing interactions based on {\hfbax}
agree well with {\hfbtho} results~\cite{hfbtho-fam}.
The coordinate-space box size is taken to be 24\,fm since this nucleus is not very weakly bound. The cutoff on the quasiparticle energies
is taken to be 80\,MeV.
In the {\hfbtho}  FAM-QRPA approach, 20 HO shells are used, with no truncations on the FAM-QRPA configuration space.
Generally, the two giant peaks, due to the deformation, are obtained in both approaches.
At the region of high excitation energy, some discrepancy between two methods can be expected.
Close to the zero energy, a spurious component appears and this is more obvious in calculations with the surface pairing (Fig.~\ref{zr100}(b)) than with the volume pairing (Fig.~\ref{zr100}(a)).
This reminds that one should be cautious when using the surface peaked pairing interaction in QRPA calculations, especially for weakly-bound nuclei
having large surface diffuseness.

\begin{figure}[t]

  \includegraphics[width=0.43\textwidth]{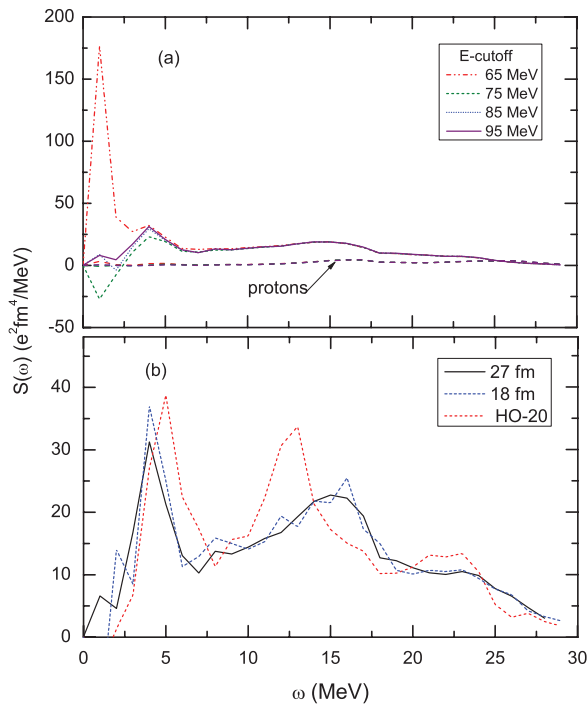}\\
  \caption{(Color online) The isoscalar monopole strength in $^{40}$Mg calculated with the
  FAM-QRPA method based on {\hfbax} and {\hfbtho}, respectively. The calculations were done with SLy4 interaction
  and surface pairing.
  (a) calculations in a coordinate-space of 27\,fm with various cutoffs on the quasiparticle energies.
  (b) calculations in a coordinate-space of 27\,fm, 18\,fm, and also on the HO basis with 20 shells. }
  \label{cutoff}
\end{figure}

In addition to $^{100}$Zr, the reliability of our approach for weakly-bound deformed nuclei has been evaluated in detail for $^{40}$Mg, by
using SLy4 interaction and surface pairing. In Fig.~\ref{cutoff}(a), the IS monopole strength of $^{40}$Mg has been calculated with
different cutoffs at quasiparticle energies. We see that at the low energy part the convergence is slow
and only with a cutoff of 95\,MeV reasonable convergence is reached.
Indeed, it is important to keep the completeness of the basis to remove the spurious contribution.
The Thomas-Fermi approximation of high energy continuum could provide a solution~\cite{pei2011}.
Fortunately, this problem is serious only for the strength at $\omega\leqslant$2\,MeV with the surface pairing.
In Fig.~\ref{cutoff}(b), $^{40}$Mg is calculated with box sizes of 18\,fm, 27\,fm and also with {\hfbtho} by using 20 shells.
As shown, all three approaches give roughly similar distributions of the transition strength. With the 18\,fm box size, the strength is rather
jagged due to less accurate continuum discretization, leading to false resonance peaks.
This stresses the importance of large coordinate-space calculations close to the neutron drip line.
By comparing the coordinate-space approach and the {\hfbtho} approach, considerable shifts in resonance peak energies can be seen,
although the agreement is good for $^{100}$Zr. This demonstrates the
necessity of the precise treatment of weak-binding effects in our approach.

\begin{figure}[t]

  \includegraphics[width=0.45\textwidth]{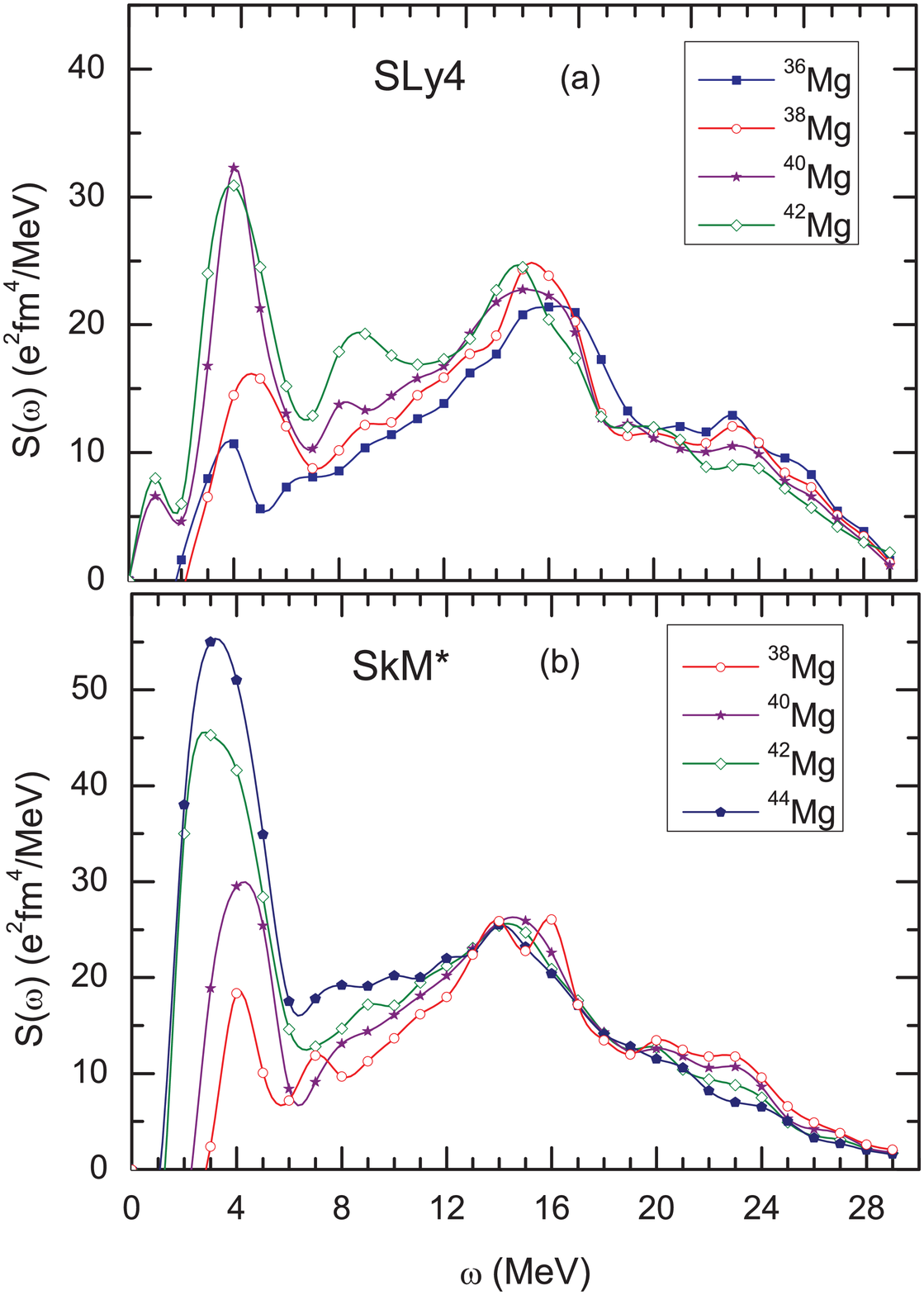}\\
  \caption{(Color online) The isoscalar monopole strength in Mg isotopes calculated with the
  FAM-QRPA method based on {\hfbax}, using the SLy4 (a) and SkM* interactions (b), respectively.}
  \label{mgiso}
\end{figure}

In Fig.~\ref{mgiso}, the calculated IS monopole strengths in Mg isotopes towards the neutron drip line are shown.
The last discovered Mg isotope is $^{40}$Mg~\cite{nature-mg}, and the drip line could be extended up to $^{46}$Mg~\cite{erler}.
The predicted drip line position is rather model and functional dependent, as shown in calculations with Skyrme functionals~\cite{erler} and
covariant density functionals~\cite{Afanasjev}.
The ground state properties and evolution of deformed halo features of $^{40,42}$Mg were studied in our previous works~\cite{pei2013}.
Fig.~\ref{mgiso} shows that low energy peaks around 4\,MeV gradually increase in both, SLy4 and SkM* calculations with the surface
pairing interaction, towards the drip line. Such kind of low energy modes were also obtained
in an earlier work which used SkM* interaction and mixed pairing~\cite{yoshida09}.
However, the peak strengths with surface pairing are significantly larger compared to those which adopted the mixed pairing, especially in calculations with SkM{*}.
Therefore, the emergence of remarkable soft monopole resonances towards the neutron drip line
is mainly related to increasing surface pairing correlations and near-threshold continuum couplings.
The two giant monopole resonances
around 15\,MeV and 23\,MeV, due to the splitting caused by deformation, are obtained in various approaches. The peak around 23\,MeV is
very weak in our calculation and it is washed out towards the neutron drip line.
$^{42}$Mg is very weakly bound in the SLy4 calculation, with a Fermi energy of $\lambda_n$=$-0.22$\,MeV.
In Fig.~\ref{mgiso}(a), a clear resonance peak around 9\,MeV appears in $^{42}$Mg and its partner in $^{40}$Mg also shows up weakly.
This monopole resonance mode is novel since it is not likely related to the deformation splitting according to the systematics.
However, this novel resonance of $^{42}$Mg is absent in SkM{*} calculations,  since the predicted stronger binding with a Fermi energy of $\lambda_n$=$-$1.19\,MeV.
In Fig.~\ref{mgiso}(b), $^{44}$Mg (with $\lambda_n$=$-$0.57\,MeV) has a considerable transition strength between 6\,MeV to 12\,MeV,
but no clear resonance peaks appear, probably due to the large strength around 4\,MeV.

\begin{figure}[tbh]
  \includegraphics[width=0.47\textwidth]{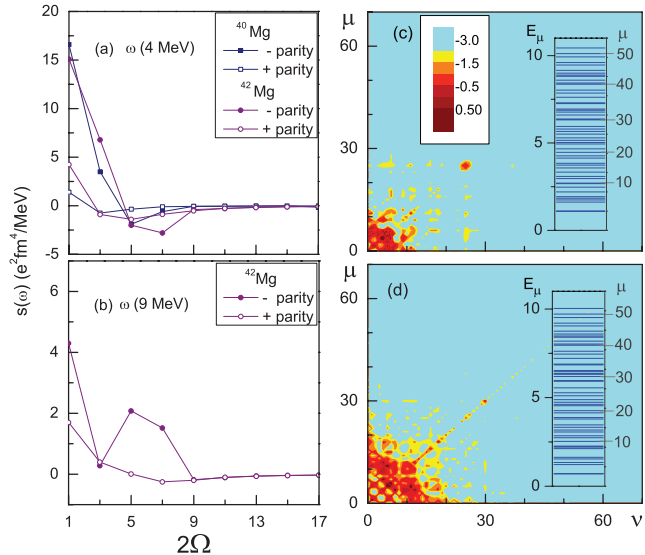}\\
  \caption{(Color online) The monopole strength from different $\Omega^{\pi}$ quasiparticle states (a, b).
  The matrix contributions of the $\Omega^{\pi}=1/2^{-}$ states corresponding to the $\omega$=4\,MeV peaks are shown for $^{40}$Mg (c) and $^{42}$Mg (d). In panels (c) and (d),
   the quasiparticle energy spectra are also displayed. See details in the text. }
  \label{collective}
\end{figure}

At the last but not the least, we have investigated the collectivity and  mechanism of the low energy monopole resonances in weakly bound Mg isotopes
with SLy4. In Fig.~\ref{collective}(a, b), the transition strengths
contributed from quasiparticle states with different values of $\Omega$ (angular momentum projection along the $z$-axis) and parity $\pi$ are shown.
As illustrated, $\Omega^{\pi}=1/2^{-}$ states contribute the majority of the strength, especially
for $^{40,42}$Mg at $\omega$=4\,MeV. This analysis leads to a conclusion that resonances around 4\,MeV region in Mg isotopes
are not very collective,
while the resonance in $^{42}$Mg around 9\,MeV is collective as shown in Fig.~\ref{collective}(b).

In Fig.~\ref{collective}(c, d), we plot a part of the matrix map of $\log(||X_{\mu\nu}|^2-|Y_{\mu\nu}|^2|)$ to
show the contributions originating from different quasiparticle states. Note that the full dimension of $1/2^{-}$ subspace is about 550.
The discretized quasiparticle energy spectra corresponding to the index $\mu$ are also displayed.
In Fig.~\ref{collective}(c), it can be seen that
the 4\,MeV resonance in $^{40}$Mg is mainly contributed by the near-threshold quasiparticle states below 2\,MeV.
Similarly, in Fig.~\ref{collective}(d), collectivity in $^{42}$Mg is enhanced which is consistent with Fig.~\ref{collective}(a).
Based on the stabilization analysis of quasiparticle spectra in $^{40,42}$Mg~\cite{pei2013}, these near-threshold $1/2^{-}$ states have a non-resonant
continuum nature and are mainly responsible for  deformed halos in weakly-bound Ne and Mg isotopes~\cite{pei2013}.
In contrast, the quasiparticle resonances  1/2$^{-}$[321] (around 3\,MeV) and 1/2$^{-}$[330] (around 6\,MeV) have no significant contributions to
deformed halos~\cite{pei2013} and to the matrix map in the case of $\omega$=4\,MeV.
The important role of non-resonant continuum in weakly bound nuclei has already been stressed (e.g., see a review paper~\cite{forssen}).
The remarkable soft monopole resonances, which are generated by non-resonant continuum, can be ascribed more likely to vibrations of
the neutron pairing halo than to vibrations of neutron matter halo. In fact, by adopting the surface pairing, the pairing density distribution has more
pronounced deformed halo features compared to neutron matter density distributions in $^{40,42}$Mg~\cite{pei2013}. This mechanism is found to
be similar to soft dipole resonances in spherical nuclei~\cite{matsuo}.
Another surprise is the second mode around 9\,MeV, related to several quasiparticle resonances, which seems to be rather collective
and has no direct relation to the non-resonant continuum by matrix analysis.  This resonance can be understood as a
novel pygmy  monopole mode due to the soft incompressibility of loosely-bound neutron skin or halo, while
the incompressibility of low density nuclear matter has not been well understood.
The experimental study of low-energy monopole modes in unstable nuclei is feasible with specially arranged detectors
around zero degrees~\cite{fayans,yeyl}.

In summary, the monopole excitations of weakly-bound deformed Mg isotopes have been studied
with the FAM-QRPA framework in large deformed coordinate-spaces, to incorporate the weak-binding effects. The reliability of our new implementation
has been carefully evaluated, to stress that the large coordinate space HFB calculations are essential close to the neutron drip line.
The systematic calculations of Mg isotopes clearly demonstrate
the emergence of collective low energy monopole modes due to two different mechanisms.
The soft monopole resonances around 4\,MeV are less collective and mainly generated by the pairing halos, corresponding to
near-threshold $\Omega^{\pi}=1/2^{-}$ non-resonant continuum states.
In addition, the strength and collectivity of such modes are enhanced towards the neutron drip line.
The emergent second resonance around 9\,MeV is rather collective and most likely linked to the expected pygmy monopole mode.
We conclude that our approach provides a suitable tool to probe novel excitation modes in weakly bound deformed nuclei, in which pairing correlations and incompressibility in low-density nuclear matter are not well understood yet.
Further developments of continuum FAM-QRPA for multipole excitations in the fully three-dimensional case will provide a broader context for questing
exotic excitation modes as well as their implications for astrophysical nucleosynthesis.

\begin{acknowledgments}
 This work was supported by the
National Key Basic Research Program of China under Grant 2013CB834400,
and the National Natural Science Foundation of China under Grants No.11375016, 11235001 and 11320101004,
and by the Research Fund for
the Doctoral Program of Higher Education of China (Grant
No. 20130001110001).
This work was also supported (M.K.) by the Academy of Finland under the Centre of Excellence
Programme 2012-2017 (Nuclear and Accelerator Based Physics Programme at JYFL) and FIDIPRO program
and by the European Union¡¯s Seventh Framework Programme ENSAR (THEXO) under Grant No. 262010.
We also acknowledge that computations in this work were performed in the Tianhe-1A supercomputer
located in the Chinese National Supercomputer Center in Tianjin.
\end{acknowledgments}

\nocite{*}


\begin{thebibliography}{999}

\bibitem{halo}
A.S. Jensen, K. Riisager, D.V. Fedorov and E. Garrido, Rev. Mod. Phys. 76,  215(2004).

\bibitem{quark}
E. S. Swanson, Phys. Rept. 429, 243(2006).

\bibitem{droplet}
S. Fantoni, R. Guardiola , J. Navarro, A. Zuker, J. Chem. Phys. 123, 054503(2005).

\bibitem{halo-exp}
I. Tanihata, J. Phys. G 22, 157(1996).

\bibitem{forssen}
C. Forssen, G. Hagen, M. Hjorth-Jensen, W. Nazarewicz, J. Rotureau, Physica Scripta T152, 014022 (2013).

\bibitem{continuum-jacek}
J. Dobaczewski, W. Nazarewicz, T. R. Werner, J. F. Berger, C. R. Chinn, and J. Decharg\'{e},
Phys. Rev. C {\bf 53}, 2809(1996).

\bibitem{Mizutori}
S. Mizutori, J. Dobaczewski, G. A. Lalazissis, W. Nazarewicz, and P.-G. Reinhard,
Phys. Rev. C 61, 044326(2000).

\bibitem{continuum}
M. Yamagami, Phys. Rev. C {\bf 72}, 064308 (2005).

\bibitem{schunck}
N. Schunck and J.L. Egido, Phys. Rev. C 78, 064305(2008).

\bibitem{matsuo2009}
{H. Oba and M. Matsuo, Phys. Rev. C {\bf 80}, 024301 (2009)}.

\bibitem{sgzhou}
S.G. Zhou, J. Meng, P. Ring, and E.G. Zhao, Phys. Rev. C 82, 011301(R)(2010).

\bibitem{pei2013}
J.C. Pei, Y.N. Zhang, F.R. Xu, Phys. Rev. C 87, 051302(R)(2013);
Y.N. Zhang, J.C. Pei, and F.R. Xu, Phys. Rev. C 88, 054305(2013).

\bibitem{paar}
N. Paar, D. Vretenar, E. Khan, G. Colo, Rept. Prog. Phys. 70, 691(2007).

\bibitem{dipole}
D. Savran, T. Aumann, A. Zilges, Prog. Part. Nucl. Phys. 70, 210(2013).

\bibitem{yoshida09}
 K. Yoshida and N. V. Giai, Phys. Rev. C 78, 064316 (2008); K. Yoshida, AIP Conf. Proc. 1165, 162 (2009).

\bibitem{reinhard}
P.-G. Reinhard and W. Nazarewicz, Phys. Rev. C 87, 014324(2013).

\bibitem{khan12}
E. Khan, J. Margueron, and I. Vida\H{n}a, Phys. Rev. Lett. 109, 092501(2012).

\bibitem{vandebrouck}
M. Vandebrouck \emph{et al.}, Phys. Rev. Lett. 113, 032504 (2014).

\bibitem{nickel}
E. Khan, N. Paar, and D. Vretenar, Phys. Rev. C 84, 051301(R)(2011).

\bibitem{hamamoto}
I. Hamamoto and H. Sagawa, Phys. Rev. C 90, 031302(R) (2014). 

\bibitem{terasaki}
J. Terasaki and J. Engel, Phys. Rev. C 82, 034326(2010).

\bibitem{fam07}
T.Nakatsukasa, T. Inakura, K. Yabana,  	Phys. Rev. C 76, 024318(2007);
T. Inakura, T. Nakatsukasa, K. Yabana,  Phys. Rev. C 80, 044301(2009).

\bibitem{fam}
P. Avogadro and T. Nakatsukasa, Phys. Rev. C 84, 014314(2011).

\bibitem{rfam}
T. Nik\v{s}i\'{c}, N. Kralj, T. Tuti\v{s}, D. Vretenar, and P. Ring  Phys. Rev. C 88, 044327(2013);
H. Liang, T. Nakatsukasa, Z. Niu, and J. Meng, Phys. Rev. C 87, 054310 (2013).

\bibitem{hfbtho-fam}
M. Stoitsov, M. Kortelainen, T. Nakatsukasa, C. Losa, and W. Nazarewicz, Phys. Rev. C 84, 041305(R) (2011).

\bibitem{nobuo}
N. Hinohara, M. Kortelainen, and W. Nazarewicz, Phys. Rev. C 87, 064309 (2013).

\bibitem{mustonen}
M. T. Mustonen, T. Shafer, Z. Zenginerler, J. Engel, Phys. Rev. C 90, 024308 (2014).

\bibitem{pei2011}
J.C. Pei, A.T. Kruppa, and W. Nazarewicz, Phys. Rev. C 84, 024311 (2011).

\bibitem{Pei08}
{J. C. Pei, M. V. Stoitsov, G. I. Fann, W. Nazarewicz, N. Schunck, and F. R.
  Xu, Phys. Rev. {\bf C78}, 064306 (2008).}

\bibitem{teran}
{E. Ter\'{a}n, V.E. Oberacker, and A.S. Umar, Phys. Rev. C {\bf 67}, 064314
  (2003)}.

\bibitem{sly4}
{ E. Chabanat, P. Bonche, P. Haensel, J. Meyer, and R. Schaeffer, Nucl. Phys. A
  {\bf 635}, 231 (1998)}.

\bibitem{skm}
{J. Bartel, P. Quentin, M. Brack, C. Guet, and H.B.
H{\aa}kansson, Nucl. Phys. A {\bf 386}, 79 (1982).}

\bibitem{mix-pairing}
{J. Dobaczewski, W. Nazarewicz, and M.V. Stoitsov, Eur. Phys. J. A {\bf 15}, 21
  (2002)}.

\bibitem{broyden}
 A. Baran, A. Bulgac, M. McNeil Forbes, G. Hagen, W. Nazarewicz, N. Schunck, and M. V. Stoitsov, Phys. Rev. C 78, 014318 (2008).

\bibitem{yoshida10}
K. Yoshida, Phys. Rev. C 82, 034324 (2010).

\bibitem{nature-mg}
T. Baumann et al., Nature 449, 1022(2007).

\bibitem{erler}
J. Erler, N. Birge, M. Kortelainen, W. Nazarewicz, E. Olsen, A.M. Perhac, M. Stoitsov,
Nature 486, 509 (2012).

\bibitem{Afanasjev}
S. E. Agbemava, A. V. Afanasjev, D. Ray, and P. Ring, Phys. Rev. C 89, 054320(2014).

\bibitem{matsuo}
M. Matsuo, K. Mizuyama, Y. Serizawa, Phys. Rev. C 71, 064326(2005).

\bibitem{yeyl}
Z.H. Yang, et al., Phys. Rev. Lett. 112, 162501(2014).

\bibitem{fayans}
S.A. Fayans, S.N. Ershov and E.F. Svinareva, Phys. Lett. B 292, 239(1992).


\end{thebibliography}

\end{document}